# An Approach to the interpretation of spontaneous transitions into a state of high conductivity in the PVC composite films

D.V.Vlasov, V.I.Kryshtob L.A.Apresyan, T.V.Vlasova, S.I. Rasmagin

General Physics Institute RAS
119991 Moscow, Russia

**ABSTRACT**
 The results of experimental studies of conductivity anomalies in film samples of specially synthesized copolymer are analyzed, in which in PVC macromolecules the polyacetylene molecular fragments are embedded with a variable concentration. Previously, it has been experimentally found that in such samples spontaneous and stimulated conductivity jumps by 12 orders of magnitude occur, and the life of each of these states can be very large - minutes, hours and more. In this paper we propose a qualitative model describing the anomalous behavior of the conductivity of the PVC composite, comprising the mechanisms of stabilization of high conductivity state, the conditions of development of instability in the transitions between the states, as well as the reasons for long living state of high conductivity even when the applied voltage is switched off. Simple numerical estimates confirming the reality of the proposed mechanisms are considered too.

## 1. Введение

В последнее время активно проводятся исследования специфики электрических свойств как проводящих (содержащих двойные сопряженные связи) [1,2], так и широкозонных полимеров [3], и композитных материалов на их основе [4], которые совместно образуют совокупность элементов с весьма широким диапазоном проводимости, от почти идеального металла до идеального изолятора. В отличие от упорядоченных проводников, полимерные материалы и композиты на их основе демонстрируют высокую специфичность проводящих свойств, в частности, радикальные измерение электропроводности при внешних воздействиях, которые открывают широкие перспективы создания новых приборов полимерной электроники типа ячеек памяти, различных датчиков и схемотехнических полимерных элементов. Некоторые элементы такой электроники, такие как органические светодиоды, уже нашли свое место на рынке, причем разработки в области литий-полимерных батарей уже превосходят по ряду параметров известные ранее типы.
 Тем не менее, на сегодняшний день многие явления и характеристики полимерных композитов мало исследованы, поскольку они имеют сложную, как правило, неупорядоченную фрактальную супрамолекулярную структуру [6,7], в которой физические процессы, такие как электропроводность и химическое строение тесно переплетены.
 В частности, к таким сложным явлениям можно отнести переключения проводящих состояний в относительно толстых пленочных образцах пластикатов и композитов ПВХ, исследованные в ряде наших публикаций [8-10]. О важности и абсолютной величине эффекта переключений можно судить на примере сополимера винилхлорида и полиацетилена [11,12], где были обнаружены спонтанные скачки проводимости на 12 порядков величины Подобные явления можно рассматривать как аналоги андерсоновского перехода металл-диэлектрик в неупорядоченных средах, или пытаться использовать другие, довольно многочисленные теории переключения проводимости неупорядоченных материалов, в частности, полимеров, включая теории возникновения пробоя.

Среди наиболее популярных моделей проводимости можно отметить многочисленные исследования полимерной проводимости за счет формирования комплексов переноса заряда [13] - полимерных донорно-акцепторных комплексов обеспечивающих проводимость полупроводникового уровня пленочных образцов. Другой подробно исследованный и описанный в монографии [14] (см. также обзор [15]) физический механизм проводимости в неупорядоченных средах – прыжковая проводимость, разработанная для слабо легированных полупроводников и возникающая за счет «прыжков» носителей заряда между атомами примеси (примесная проводимость и моттовский переход метал-диэлектрик), которая отчасти применима для описания некоторых случаев переключения полимерной проводимости. При этом «перескоки» могут быть связаны как с термической активизацией («некогерентный» надбарьерный механизм [16]), так и с подбарьерным туннелированием («когерентный» квантовый механизм [17]). Известны также многочисленные теории механизмов проводимости в тонких нанометровых пленках, где на первое место выступают контактные эффекты, связанные с инжекцией носителей и возникновением токов, ограниченных пространственным зарядом [18,19], а также описанные в обзоре [3] различные механизмы переключений проводимости, наблюдаемые в тонких пленках широкозонных полимеров.

Обычно аномально высокая проводимость пленок аморфных полупроводников связана с возникновением тонких проводящих каналов, расположенных перпендикулярно поверхности [20-23]. При этом во многих исследованиях специально отмечалось существование граничной критической толщины $L_{кр}$ при превышении которой, скачки проводимости не наблюдались, причем обычно наблюдаемая толщина $L_{кр}$ ограничивается субмикронным или даже нанометровым диапазоном [3]. В отличие от этого, в наших работах исследовались относительно толстые пленки, с толщиной от десятков до сотен микрон.

Тем не менее, на наш взгляд, существующие теоретические модели по большей части построены ad hoc и не описывают наблюдаемые нами аномальные явления спонтанных переключений проводимости, устойчивость и значительную продолжительность времени жизни СВП, относительно быстрое переключение состояний проводимости в полимерных композитах на базе ПВХ.

## 2. Совокупность новых экспериментальных данных по переходу композитных материалов ПВХ в СВП

В результате исследований проводимости полимерных композитов [11,12] нами было обнаружено, что в образцах композита, приготовленных с использованием частично дегидрохлорированного ПВХ, проявляются аномальные эффекты, связанные в основном с гигантскими скачкообразными переключениями состояний проводимости, которые составляли до 12 порядков величины: от уровня сопротивления изолятора порядка $10^{12}$ Ом до единиц и долей Ома.

С другой стороны, химическая структура макромолекул дегидрохлорированного ПВХ хорошо известна, поскольку дегидрохлорирование фактически эквивалентно одному из известных механизмов термического старения изоляционных материалов [24]. При термической обработке раствора ПВХ происходит отщепление молекулы HCl от макромолекулы, в результате чего образуется звено цепи макромолекулы содержащее двойную связь — C=C— . При продолжении процесса выделения HCl в макромолекулы ПВХ встраиваются цепочки двойных сопряженных связей типа: ….. —C=C—C=C—…… случайной длины, которые имеют относительно слабо связанные π-электроны и фактически могут при определеннных условиях превращаться в нано-проводники.

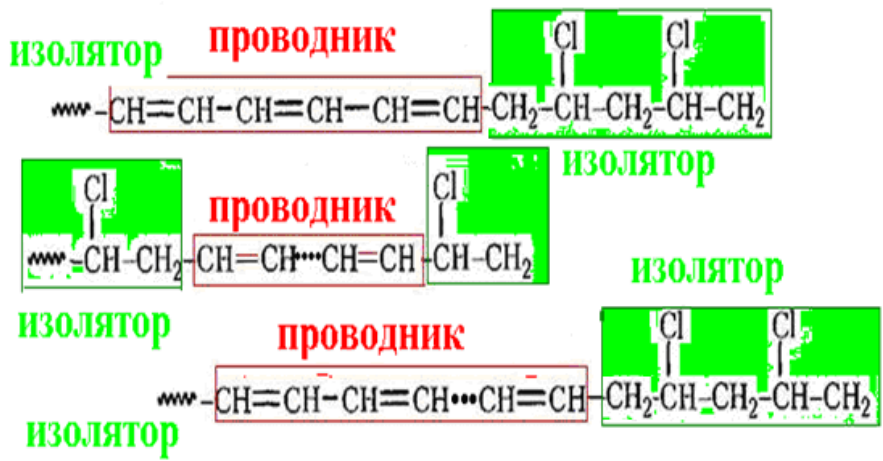

Рис.1. Структура макромолекул специально синтезированного полимера (в реальных образцах макромолекулы ориентированы хаотически).

Используя описанный в [11,12] процесс, нами был синтезирован Сополимер, содержащий Сопряженные двойные связи (СС) на базе ПВХ, в макромолекулы которого встраиваются фрагменты молекул проводящего полимера - полиацетилена (ПАц, Рис.1). В зависимости от степени дегидрохлорирования, на начальной стадии процесса материал является классическим изолятором - широкозонным полимером. По мере увеличения концентрации цепочек двойных сопряженных связей пленочные образцы, полученные из такого термолизованного ПВХ, моделируют композит с проводящими включениями и изолирующими промежутками.

Отметим также, что добавление фрагментов двойных сопряженных связей в процессе дегидрохлорирования ПВХ приводило к появлению в пленках, полученных методом полива, определенных пластических свойств, так что не было необходимости в использовании каких либо пластификаторов для изготовления достаточно эластичных для экспериментальных измерений пленочных образцов (от 10 до 200 мкм). В наших экспериментах по исследованию электропроводности [12] такие сополимеры продемонстрировали самый высокий уровень проводимости в СВП по сравнению с другими исследованными ранее композитами ПВХ, в которых начальный уровень проводимости повышался введением специального пластификатора-модификатора А [8-10].

На Рис.2 приведена (в логарифмическом масштабе) типичная зависимость сопротивления пленочных образцов СС от времени термолиза, приведенная в работе [12] в виде таблицы. Этот график показывает, что сопротивление образцов начиная с некоторого уровня дегидрохлорирования обнаруживает склонность к спонтанным переходам в СВП с изменением сопротивления на 10 и более порядков величины. Под термином «спонтанный» здесь понимается переход в СВП «без видимых внешних воздействий», т.е. при сохранении значений приложенного напряжения, температуры, давлении, освещении и т.д.

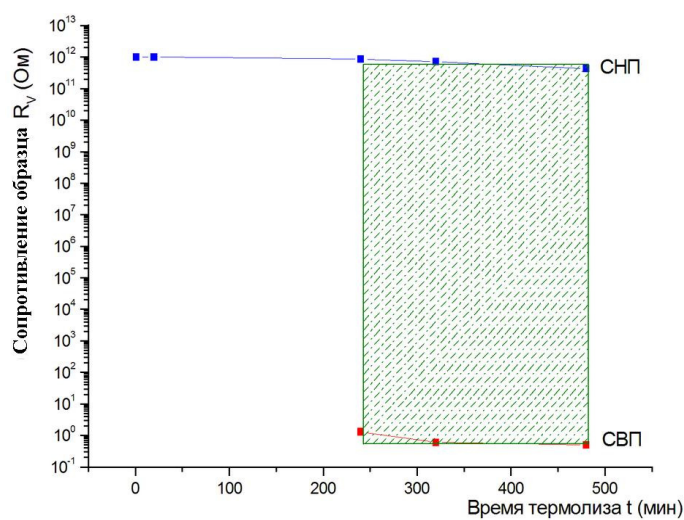

Рис.2. Типичная зависимость сопротивления образца СС от степени дегидрохлорирования раствора ПВХ при температуре $190^0$ С. Заштрихована зона неустойчивости относительно перехода между состояниями высокой (СВП) и низкой (СНП) проводимости: СВП ←→ СНП.

Высокий перепад сопротивлений и хорошая воспроизводимость режимов спонтанных переходов, а также известная из литературы [24] химическая структура материала, получаемого при дегидрохлорировании ПВХ дают возможность, по крайней мере на качественном уровне, проанализировать физические механизмы наблюдаемых спонтанных скачков проводимости. Это позволяет строить качественные модели, описывающие наблюдаемые явления, в том числе наличие двух устойчивых состояний проводимости (точнее, относительную устойчивость состояния высокой проводимости (СВП)), а также широкий спектр наблюдаемых времен жизни СВП (от секунд до часов и суток).

Полимерная специфика гигантских скачков проводимости в СС образцах проявляется в том, что измеряемая проводимость образцов СС «через» пленку падает до долей Ома, но в то же время попытка измерить сопротивление вдоль пленки в ситуации, когда заведомо реализуется СВП, продемонстрировала отсутствие изменений поверхностного сопротивления пленки $R_s$ ( более $10^{12}$ Ом), аналогично тому что наблюдалось в пластифицированных ПВХ композитах [8-10]. При этом справедливое в случае однородного распределения токов выражение для сопротивления R = ρl/S, где ρ, l и S удельное объемное сопротивление, длина и поперечное сечение образца, соответственно, при разумном выборе параметров дает для $R_s$ оценку порядка 10 кОм, что очевидным образом не соответствует экспериментальным данным (в эксперименте > $10^{12}$ Ом ) и свидетельствует в пользу сильной неоднородности распределения токов в образце. Фактически, в отличие от «нормальной» проводимости, наблюдаемые в СС скачки проводимости одномерны и направлены перпендикулярно пленке, т.е. возникающее СВП имеет сильную анизотропию (более $10^9$ ) на молекулярном уровне – направленные микро каналы проводимости.

Другой особенностью скачков проводимости в рассматриваемых здесь ПВХ композитах можно считать широкий разброс продолжительности времен жизни СВП. Так, на Рис.3. показана одна из реализаций случайного, спонтанного перехода из СНП в СВП, и, после некоторого периода (около десяти секунд) соответствующего обратного перехода в СНП. Отчетливо видно, что время жизни СВП существенно превышает времена переходов между состояниями.

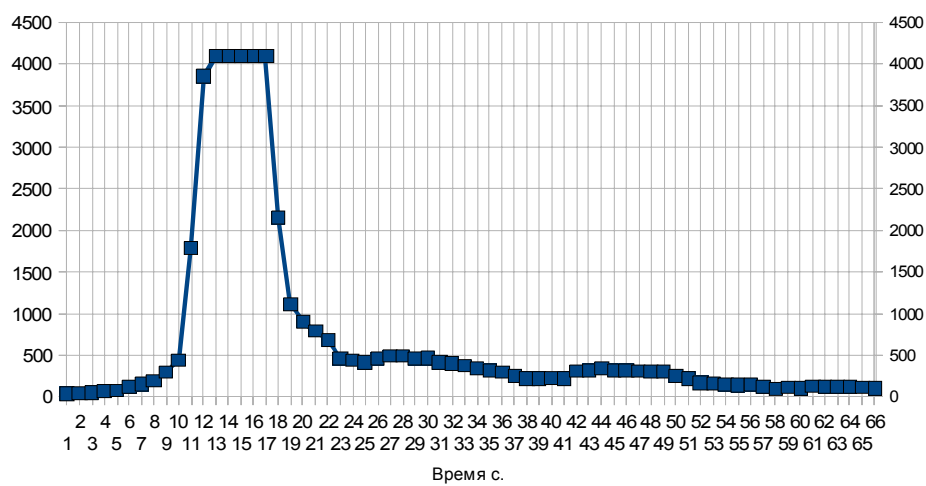

Рис.3 Осциллограмма спонтанного перехода в СВП образца СС-полимера и его самопроизвольный возврат в СНП [26]. Плоская вершина – ограничения АЦП.

## 3.Анализ физического механизма аномальной проводимости

В типичных схемах измерения проводимости изоляторов с большим балластным сопротивлением переход в СВП при увеличении напряжения реализуется всегда, в частности в наших экспериментах [8-10] такой обратимый переход, называемый нами мягкий «пробой», наблюдался на всех ПВХ образцах при напряжениях порядка 30 В/мкм, и на всех других полимерных образцах (полиэтилен, тефлон и т.д.). Этот эффект универсален и не зависит от типа полимера, и описывается многочисленными теориями пробоя диэлектриков [25], т.е. высвобождением электронов из примесных атомов, мелких ловушек и дефектных узлов макромолекул, набором энергии во внешнем поле с дальнейшей ионизацией молекул основной матрицы. Термин «мягкий» пробой мы применили к процессу пробоя, когда ток, возникающий при пробое существенно ограничен балластным сопротивлением и энергия пробоя недостаточна для изменения структуры материала или образования необратимых изменений молекулярной структуры. Поэтому «мягкий» пробой является неразрушающим и обратимым, в отличие от реального пробоя, где происходят необратимые изменения структуры образца за счет выделяемого тепла и искрового разряда.

Все рассматриваемые в наших статьях экспериментальные результаты по скачкам проводимости были получены при напряжениях, как минимум на порядок ниже порогов мягкого пробоя, причем в данной работе мы обсуждаем лишь спонтанные переходы, в то время как исследуемые обычно стимулированные переходы происходят под влиянием контролируемых внешних воздействий.

Для композитного материала типа СС, состоящего из проводящих элементов (нано- или микро- частиц), разделенных изолирующими промежутками, проводимость так или иначе связана либо с туннелированием, либо с перескоками электронов между проводящими элементами композита. В отличие от «классической» прыжковой проводимости, в которой носители заряда «перескакивают» с атома на атом, в случае композита такие перескоки имеют место между проводящими элементами, тогда как вдоль проводящего элемента заряды распространяются свободно. При этом возникает первое отличие, на которое следует обратить внимание в рассматриваемом случае, а именно, существенное эффективное увеличение локальных значений электрического поля за счет введения проводящих элементов (Рис.4).

Если, следуя [26], использовать упрощения выражения для вероятности W туннелирования между двумя проводящими элементами, разделенными промежутком $\Delta x$, то можно записать:

$$W(\Delta x, U_0) = W_0 \exp(-\Delta x / l_d), \qquad (1)$$

где $1/l_d = \sqrt{8mU_0}/\hbar$, $l_d$ - характеристическая длина, определяемая потенциальным барьером $U_0$ и эффективной массой заряда m. Если расстояние между электродами равно L, то можно ввести эффективную суммарную длину проводящих элементов на траектории прохождения заряда от одного электрода до другого $L_C$, и тогда разность $L-L_C$ будет по порядку величины соответствовать эффективной длине изолирующих промежутков. Для получения приблизительной оценки вероятности туннелирования заряда с одного электрода на другой $W_{12}$ путем статистически независимых перескоков по одной из возможных траекторий между электродами, запишем выражение для полной вероятности в качестве произведения вероятностей для каждого отдельного промежутка:

$$W_{12} = \prod W_i = W \exp\{-(L-L_c)/l_d\}. \qquad (2)$$

Полный ток будет определяться суммой всех вероятностей (2) по всем траекториям, однако очевидно, что экспоненциальный множитель, в каждом слагаемом сократит сумму до нескольких членов с минимально короткими (суммарно) изолирующими промежутками. Таким образом, исходя из (3) можно считать, что ток будет протекать лишь в нескольких наиболее «коротких» каналах и при определенных условиях может приводить к экспоненциальной зависимости проводимости от длины образца [10].

С другой стороны, из выражения (2) следует, что при прохождении порога перколяции, т.е. при после «замыкании» цепочки проводящих примесей ток через образец композитного материала будет определяться проводимостью проводящих элементов, при этом должен наблюдаться скачек тока на много порядков величины. Очевидно, что при приближении композитного материала к порогу перколяции за счет увеличения концентрации проводящих примесей вероятность туннелирования и, соответственно, ток будут резко увеличиваться и именно в этой области можно ожидать необычного поведения проводимости. Действительно, можно показать, что именно вблизи порога перколяции композитная среда может быть неустойчивой относительно перехода в СВП, причем конкретные механизмы такого перехода, кроме туннелирования и перескоков могут, вообще говоря, дополняться специфическими для данного композита механизмами переноса заряда.

Для наглядности рассмотрим структуру поля в композитном образце приведенную на Рис.4. Для однородного полимерного образца напряженность поля постоянна и его величина определяется как $E_0 = U_0/L$. При добавлении проводящих элементов эффективное среднее электрическое поле «скапливается» в изолирующих промежутках между проводниками:

$$E = \frac{U_0}{L - L_0} \qquad (3)$$

и его значение будет увеличиваться по мере увеличения эффективной суммарной длины проводящих элементов.

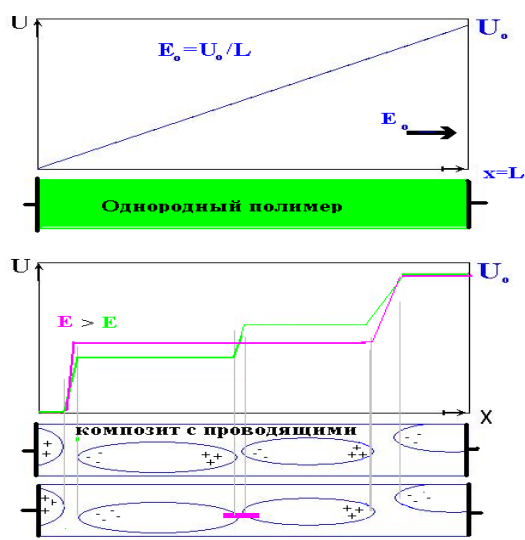

Рис.4. Сравнение структуры поля в однородном (верхний график) и композитном материалах (рисунок снизу).

В описываемых экспериментах с различными концентрациями проводящих элементов, даже заведомо ниже порога перколяции наблюдался заметный ток. Поэтому, если в композите может быть реализован какой-либо физический механизм нелинейности, который может «замкнуть» два соседних проводящих элемента (например, всегда есть вероятность мягкого пробоя), то поле в соседних изолирующих промежутках увеличится, и уже в следующем промежутке может произойти соответствующий переход в проводящее состояние и т.д.  Таким образом, получаем, что композитная среда обладает неустойчивостью перехода в проводящее состояние, причем эта неустойчивость тем сильнее, чем ближе заполнение полимера проводящими элементами к порогу перколяции.

Наличие неустойчивости каскадного замыкания изолирующих промежутков объясняет часть наблюдаемых результатов, т.е. относительно быстрый переход, в том числе спонтанный,  в СВП  за  время  существенно более короткое, чем времена жизни собственно СВП и СНП. Тем не менее, остается вопрос,  каков механизм устойчивости СВП. Для качественного ответа на этот вопрос обратимся к Рис.5, который в рамках предлагаемой гипотезы поясняет, каким образом в композитных материалах возникает устойчивость СВП. В соответствии с развиваемой моделью,  в СВП все перемычки между проводящими элементами замкнуты (например, вследствие мягкого пробоя). Если теперь (например, вследствие тепловых флуктуаций) один из элементов размыкается, то все электрическое поле скапливается именно в этом промежутке- изоляторе, который с большой вероятностью вновь  испытывает мягкий пробой и становиться проводящим.

Рассматриваемый каскадный механизм перескока в СВП универсален, т.е. не зависит от материала диэлектрика и механизма замыкания проводящих включений. Тем не менее, очевидно, что переход в СВП связан с вольтамперной характеристикой, и в случае спонтанных переходов, такая характеристика должна быть функцией случайных параметров, а главное, должна иметь механизм памяти о предшествующем состоянии образца, что может приводить к гистерезису. Отметим, что при перескоках в СВП за счет внешне приложенного поля  в отсутствие гистерезиса  при снятии напряжения с образца СВП должно исчезать.

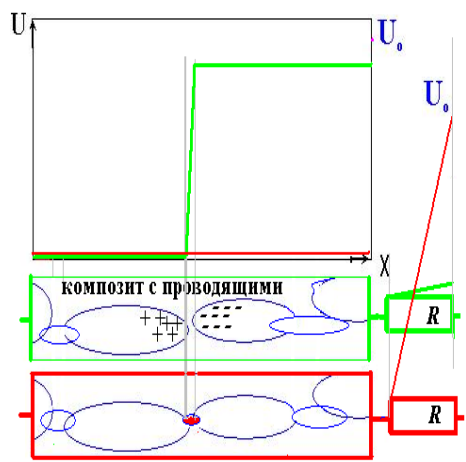

Рис. 5. Устойчивость проводящего состояния в композитном материале. В нижней  части рисунка показана диаграмма распределения потенциала в СВП образца  (все напряжение падает на балластном сопротивлении, см. справа);  при флуктуационном переходе одного из проводящих элементов в СНП все напряжение падает на изолирующем мостике.

Таким образом, наличие конечного и иногда достаточно продолжительного времени жизни и относительная устойчивость СВП за счет каскадного механизма слегка проясняют ситуацию с перескоками. Остается необъясненным тот экспериментальный факт, что образец ПВХ композита может сохранять состояние проводимости даже при снятии внешнего поля, в том числе на достаточно длительных временных интервалах порядка минут, часов и более [11,12]. Для объяснения такого поведения, можно снова привлечь процесс дегидрохлорирования, т.е. происходящего при достаточно существенном нагревании (порядка $100^0$ С) отделения от макромолекулы ПВХ атомов водорода и хлора с образованием в макромолекуле двойной связи, что способствует резкому увеличению проводимости.

В отличие от термолиза за счет внешнего нагрева, о котором шла речь выше в Разд.2, в данном случае нагревание может происходить в результате выделения джоулева тепла непосредственно в толще образца при прохождении тока через проводящий канал. При этом будет эффективно увеличиваться концентрация двойных связей в области изолирующего промежутка и, соответственно, уменьшается его сопротивление т.е. возникает замыкание. Важно, что если этот процесс дегидрохлорирования происходит внутри твердого образца ПВХ, то движение «квазимолекулы» HCl, отделившейся от макромолекулы полимера, ограниченно свободным объемом с размерами порядка нанометров (см. Рис.6). Находясь в непосредственной близости от двойной связи, такая «квазимолекула» может вновь войти в структуру макромолекулы ПВХ, превращая ее в изолятор. Такой механизм обратимого локального дегидрохлорирования может, в частности, порождать гистерезис, связанный с химической задержкой восстановления проводимости. Такой процесс может обеспечить широкий разброс характерных времен релаксации СВП и, в то же время, сохраняет обратимость переходов между состояниями проводимости.

Рис.6 Срез полимерной структуры с указанием свободного объема в виде наноразмерных неупорядоченных каверн, с указанием «квазимолекулы» HCl, запертой в одной из каверн. Стрелкой указан путь атома хлора от места своего закрепления. При случайном блуждании в каверне свободного объема атом Cl может вернуться на свое место и заблокировать двойную связь.

Наличие проводящих каналов при переходе в СВП в образцах СС наблюдалось нами экспериментально при комнатной температуре (24 $^o$ С). При этом использовался специальный прозрачный электрод (слой ITO на стеклянной подложке) с проводимостью 100 Ом на квадрат. При переходе в СВП образца СС (полученного дегидрохлорированием в течение 480 минут) через прозрачный электрод можно было при специальном затемнении наблюдать несколько слабо светящихся точек, которые на наш взгляд связаны с узкими токоведущими каналами и аналогичны наблюдениям, описанным и в других публикациях по исследованию электропроводности полимерных пленок [20-23]. Интенсивность светящихся точек была различной и изменялась со временем, причем в некоторых каналах развивалось кратковременное до 10 секунд интенсивное свечение, величина которого была достаточной для уверенной регистрацией фотоаппаратом. Характерная «вспышка» канала показана на Рис.7 .

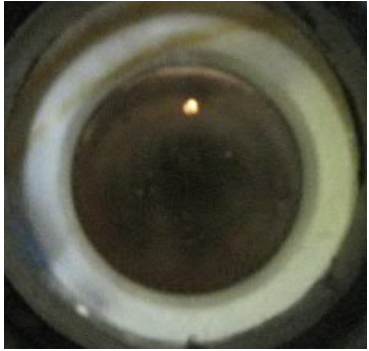

Рис.7. Наблюдение вспышки светового излучения от токоведущего канала, возникающего в пленке образца СС при переходе в СВП. Размер кольца затемнения образца 40 мм. Оценка толщины канала, требует специальных измерений, поскольку в данном случае использована камера с автофокусом.

Для получения грубой численной оценки локального разогрева полимерной пленки толщиной 10 мкм в условиях, близких к экспериментальным [11,12] подсчитаем разогрев вещества в таком токоведущем канале. Будем считать что полная мощность источника, нагрева токоведущий канал, диссипирует в объеме ПВХ за счет теплопроводности через стенки канала за время $t \sim S_\perp/\chi$, где $S_\perp$ -площадь поперечного сечения канала, а $\chi$ - коэффициент температуропроводности, связанный с теплопроводностью к, плотностью $\rho_{ПВХ}$ и теплоемкостью $C_{ПВХ}$ известным соотношением $\chi = к/(\rho_{ПВХ} C_{ПВХ})$. В результате для температуры нагрева канала получаем оценку

$$\Delta T \sim (U I)/(\rho_{ПВХ} C_{ПВХ} V) \; (S_\perp/\chi) = (U I)/(к L) \qquad (4)$$

Подставляя сюда U=10V, I= $10^{-5}$A, к = 0.15 W/(m K), L=$10^{-5}$ m, получаем оценку изменения температуры канала $\Delta T \sim 67°$. Таким образом, по порядку величины мы попадаем в диапазон температур начала активного дегидрохлорирования ПВХ (от 70 до 200°C). Резкое уменьшение сопротивления токоведущего канала может привести к соответствующему увеличению мощности, возбуждению электро-люминесценции, которые можно наблюдать экспериментально.

Разумеется, наиболее критическим для данной оценки является использование макроскопических т.е.усредненных характеристик ПВХ, тогда как реально развитие неустойчивостей происходит на малых масштабах, где среда является сильно неоднородной, что должно приводить к сильным флуктуациям напряженности поля, проводимости а также геометрических характеристик канала. Как следствие, выделение тепла в канале должно иметь неоднородный характер с максимумами в областях наименьшей проводимости. С другой стороны, и напряжения, прикладываемые к образцу, и ток также изменялись в экспериментах в достаточно широких пределах, так что порядок оценки изменения температуры в канале может попадать в диапазон локального развития реакции дегидрохлорирования

## 4. Заключение

В данной работе мы развили качественную модель наблюдаемых в ПВХ-композитах перескоков, позволяющую объяснить обнаруженные ранее в пленочных образцах (толщиной более 10 мкм) частично дегидрохлорированного ПВХ обратимые переходы из состояния низкой проводимости в состояние высокой проводимости с амплитудой скачков до 12 порядков. С этой целью расширена предложенная ранее в [26] универсальная модель каскадного перехода композитного материала в СВП и указаны механизмы неустойчивости, обеспечивающей относительно быстрые переходы образца из СНП в СВП и обратно.

Для ПВХ–композитов (и ему подобных полимеров, например, поливинилбромида) указан более низкопороговый по сравнению с мягким пробоем механизм перехода в СВП в рамках той же каскадной модели. Предложен структурно-химический механизм, объясняющий относительно длительные времена жизни СВП в ПВХ композитах, и основанный на том, что вблизи порога перколяции происходит заметный локальный разогрев ПВХ изолирующих промежутков, активизирующий (за счет изолированного элемента свободного объема) обратимую реакцию дегидрохлорирования с отделением «квазимолекулы» $HCl$, движение которой ограничено свободным объемом. С практической точки зрения с учетом развиваемой модели более тщательное исследование распределения времен жизни СВП может быть использовано для изучения свойств и структуры свободного объема в ПВХ-композитах, наряду или вместо достаточно дорогого и сложного электрон-позитронного метода [27,28].